\documentclass[prb,twocolumn,amsmath,floats,showpacs]{revtex4}

\usepackage{graphicx}
\usepackage{epsfig}

\begin{document}
\input epsf

\title{Investigation of the superconducting energy gap in the compound
LuNi$_{2}$B$_{2}$C by the method of point contact spectroscopy:
 two-gap approximation}

\author{N.~L.~Bobrov, S.~I.~Beloborod'ko, L.~V.~Tyutrina, V.~N.~Chernobay,
I.~K.~Yanson \footnote {Email address: yanson@ilt.kharkov.ua} }
\affiliation{B.I.~Verkin Institute for Low Temperature Physics and
Engineering, NAS of Ukraine, 47, Lenin Prospect, 61103, Kharkiv,
Ukraine}

\author {D.~G.~Naugle and K.~D.~D.~Rathnayaka}
\affiliation{Department of Physics Texas A\&M University, College
Station TX 77843-4242, USA}

\date{\today}

\begin{abstract}
It is shown that the two-gap approximation is applicable for
describing the $dV/dI(V)$ spectra of LuNi$_{2}$B$_{2}$C-Ag point
contacts in a wide interval of temperatures. The values and the
temperature dependences of the large and the small gaps in the $ab$
plane and in the $c$ direction were estimated using the generalized
BTK model~\cite{Plecenik} and the equations of.~\cite{Beloborodko}
In the BCS extrapolation the critical temperature of the small gap
is $10$ $K$ in the $ab$ plane and $14.5$ $K$ in the $c$ direction.
The absolute values of the gaps are $\Delta_0^{ab}=2.16$ $meV$ and
$\Delta_0^c=1.94$ $meV$. For the large gaps the critical temperature
coincides with the bulk $T_c$, $T_c^{bulk}=16.8$ $K$, and their
absolute values are very close, being about $3$ $meV$ in both
orientations. In the $c$ direction the contributions to the
conductivity from the small and the large gaps remain practically
identical up to $10 \div 11$ $K$. In the $ab$ plane the contribution
from the small gap is much smaller and decreases rapidly as a
temperature rises.

\end{abstract}

\maketitle

\section{Introduction}
Compounds {\it Re}Ni$_{2}$B$_{2}$C {\it Re} is a rare earth metal)
have a tetrahedral crystal structure (e.g., see~\cite{Muller}
Fig.\,6) with the spatial group $I4/mmm$~\cite{Muller} and the
characteristic parameters: two lattice constants $a$ and $c$
($a=3.46$ \AA, $c=10.63$ {\AA} for LuNi$_{2}$B$_{2}$C) and the
distance $z$ between the $B$ atom and the $RC$ plane. The anisotropy
in these compounds is determined by the ratio $c/a$ and is rather
high ($c/a \sim 3$).~\cite{Muller}

According to electron structure calculation, {\it
Re}Ni$_{2}$B$_{2}$C compounds are three-dimensional metals in which
all atoms contribute to the density of states at the Fermi
surface.~\cite{Kim,Mattheiss,Dugdale} In LuNi$_{2}$B$_{2}$C three
energy bands crossing the Fermi level subdivide the Fermi surface
into three distinct regions: (1) two spheroidal sheets at
$\Gamma$-points that are extend along the $c$ axis; (2) two flat
P-centered square regions with the sides parallel to
$\langle100\rangle$ and $\langle010\rangle$, and finally (3) the
largest basic $X$-centered part mostly cylindrical in shape (closely
resembling practically two-dimensional crystal structure) whose axis
is parallel to the $c$ axis.~\cite{Kim,Mattheiss,Dugdale} In this
sheet there are flat regions at the points $0.56{\pm}(2{\pi}/a)$,
the nesting, typical for the compounds of this
family.~\cite{Dugdale} It is precisely these features that determine
the Kohn anomaly near the wave vector
$Q_{m}\approx(0.5,0,0)$~\cite{Dervenagas,Stassis} and the
incommensurate ordering with $Q_{m}\approx(0.5,0,0)$ in AFM
compounds. The part of the Fermi surface responsible for nesting
makes only $(4.4\pm0.5\%)$ and shows up a slightly increased
resistance for the current along $\langle100\rangle$.~\cite{Dugdale}

As was shown for point contacts with a ballistic transit of
electrons in a many-band superconductor~\cite{Mukhopadhyay}, in any
direction the contribution to conductivity from each band is
proportional to the area of the Fermi surface projection from the
corresponding band onto the interface. Since the fraction of the
Fermi surface responsible for nesting is small, the nesting-related
features are not obvious in the point-contact spectra.

Thus, the Fermi surface of nickel borocarbides consists of several
bands: it is anisotropic and corresponding Fermi velocity $v_F$
varies within wide limits. In this case $T_c$ is determined not by
the total density of states $N(E_{F})$, but by the narrow peak of
the density of states at the Fermi level (see~\cite{Muller},
Fig.\,12). The density-of-states peak is induced by the slow
electrons from the flat (nesting) regions of the Fermi surface with
the wave vectors (0.5-0.6,0,0).~\cite{Muller}

In the normal state the resistivity $\rho$ is isotropic because it
is connected with the groups of electrons that have rather high
velocities $v_F$ and do not belong to the flat regions of the Fermi
surface.~\cite{Muller} The temperature dependence of resistance
$\rho$ in single crystal YNi$_{2}$B$_{2}$C and LuNi$_{2}$B$_{2}$C
also exhibits this practically isotropic metallic behavior
(see~\cite{Muller}, Fig.\,11).

The scanning tunnel microscope measurements on LuNi$_{2}$B$_{2}$C
give a rather low value of superconducting energy gap along the
$c$ axis (4.2\, K),
 $\sim 2.2$ $meV$ ($2\Delta/{kT_c}=3.2$).~\cite{DeWilde,DeWilde1} In the
$ab$ plane ($T\sim 0.3 \div 0.5$ $K$) for YNi$_{2}$B$_{2}$C the gap
data are greatly scattered, from $2.3$ \,{$meV$}~\cite{Martinez} to
$3.5$ \,{$meV$}~\cite{Nishimori} ($2\Delta/kT_c=3.44 \div 5.24$).
The high ratio in~\cite{Nishimori} can be attributed to the
estimation procedure: the gap was estimated from the positions of
the maxima in the differential conductivity curve. The shape of the
curve, however, suggests a considerable variation of the gap over
the Fermi surface, and the cited value of the gap is closer to its
maximum rather than to its average magnitude. For a number of nickel
borocarbides the ratio obtained from point contact measurements in
the $ab$ plane~\cite{Yanson,Rybalchenko} is $2\Delta/kT_c=3.7 \div
3.8$.

In the recently discovered superconductor MgB$_{2}$ the Fermi
surface is formed by two groups of energy bands having different
dimensions, which determines a two-gap conductivity~\cite{Yanson1}.
Similarly, we can expect a many-gap conductivity in nickel
borocarbide compounds. Note that the dHvA data~\cite{Goll,Nguyen}
also point to the presence of at least two or even three groups of
the Fermi velocities evidencing in favor of a two or three-band
model. In~\cite{Shulga} the behavior of $H_{C2}(T)$ is considered
within a two-band model. One band has low Fermi velocities, high
$H_{c2}(0)$, $T_c$, $\lambda$ (electron-phonon interaction)
parameters. However, all these characteristics are suppressed in the
bound system by another band with high Fermi velocities and smaller
superconducting parameters. In this case the two-gap approximation
commonly used for two-band superconductors~\cite{Takasaki} seems
quite reasonable.

Here we report the results on energy gap anisotropy that were
obtained on the nickel borocarbide superconductor LuNi$_{2}$B$_{2}$C
for the main crystallographic orientations - in the $ab$ plane and
along $c$ axis. This study is a logical extension of~\cite{Bobrov}.
It was shown that the one-gap approximation could not provide an
accurate description of the experimental curves in the low
temperatures region and use of the two-gap approach was
contemplated.~\cite{Bobrov} The largest and the smallest gaps were
estimated in the $ab$ plane and along $c$ direction at $T \ll T_c$.
Besides, using the generalized BTK~\cite{Plecenik} and
Beloborodko`s~\cite{Beloborodko} models, the lowest values of the
largest gap were estimated in these directions as a function of
temperature.~\cite{Bobrov} Owing to the high quality of the
experimental curves (no "humps", or broad maxima at $|eV| >
\Delta$), the estimation was possible in the "wings", i.e. in the
regions of the experimental curves where the biases are higher than
that at the differential resistance minima ($|eV| \approx \Delta$).

Below we show the results of the two-band calculation for
LuNi$_{2}$B$_{2}$C obtained within two
models~\cite{Plecenik,Beloborodko} for experimental curves obtained
before.~\cite{Bobrov} In our opinion, the calculation most closely
corresponds to the many-band character of the Fermi surface in these
compounds. The anisotropy is estimated qualitatively in each band
using the parameter $\Gamma$ in the BTK model,~\cite{Plecenik} or
$\gamma$ in Beloborodko's model.~\cite{Beloborodko} The calculation
is compared with the results of one-gap fitting.~\cite{Bobrov}

\section{Experimental technique}

The point-contact measurement was performed on single crystal
LuNi$_{2}$B$_{2}$C ($T_{c}\simeq16.9$ $K$) grown by Canfield and
Budko using a flux method.~\cite{Cho} The crystals were thin ($0.1
\sim 0.2$ mm) plates with the $c$ axis perpendicular to the plane of
the plate. The single crystal surface always has quite a thick layer
on it in which superconductivity is either absent or strongly
suppressed. Point contacts between natural crystal faces therefore
suffer from this disadvantage. For measurement in the $ab$ plane,
the crystal is usually cleaved, and the point contact is made
between a metallic counterelectrode and the cleaved surface. The
cleavage perpendicular to the $c$ direction is a technical
challenge. To do this, the crystal surface was cleaned with a
$10\div15\%$ $HNO_3$ solution in ethanol for several minutes. The
properly etched crystal has metallic luster and its surface is free
from colored film. The crystal was washed thoroughly in pure ethanol
and than installed in contact-making device. As found subsequently,
the etching conditions are very important for producing high-quality
point-contacts.~\cite{Naidyuk} In the $ab$ plane the measurements
results seem to be indifferent to cleavage or etching - they are
practically similar in both cases. But a deeper analysis (see below)
reveals certain distinctions caused by defects in the cleavage
surface. The other electrode was made of high-purity silver. The
point contacts in the $ab$ plane were fabricated between the edge of
the prism-shaped $Ag$ electrode and freshly cleaved (etched) face of
the single crystal (shear technique~\cite{Chubov}). The deviation
from the direction perpendicular to the $c$ axis varied within
$5\div10^\circ$. To made a contact in the $c$ direction, the
"needle-anvil" geometry was used. The needle radius was 1-3 microns.
The temperature was measured with a special insert (similar to that
in~\cite{Engel}).

The point contact resistance varied typically from several Ohms to
tens of Ohms. To attain more detailed data, we selected
point-contacts with the greatest possible tunneling which was judged
from the presence of a differential resistance maximum at zero bias
and the highest nonlinearity of the $I$-$V$ curves at biases $\pm
\Delta$ typical for a nonperturbed superconducting surface in the
contact neighborhood. Unfortunately, we were able to take a complete
set of curves in the whole range $T_{min} \approx 1.5 K \rightarrow
T_c$ only on a few contacts. Because of high resistance and
long-term (over 10-12 hours) measurements of temperature, many of
contacts were broken down. The measurements were made with the
nearly equal temperature steps. There were taken 47 curves in the
$c$ direction and 41 curves in the $ab$ plane.

\section{Data processing}

\begin{figure}
\centering
\includegraphics[width=\columnwidth]{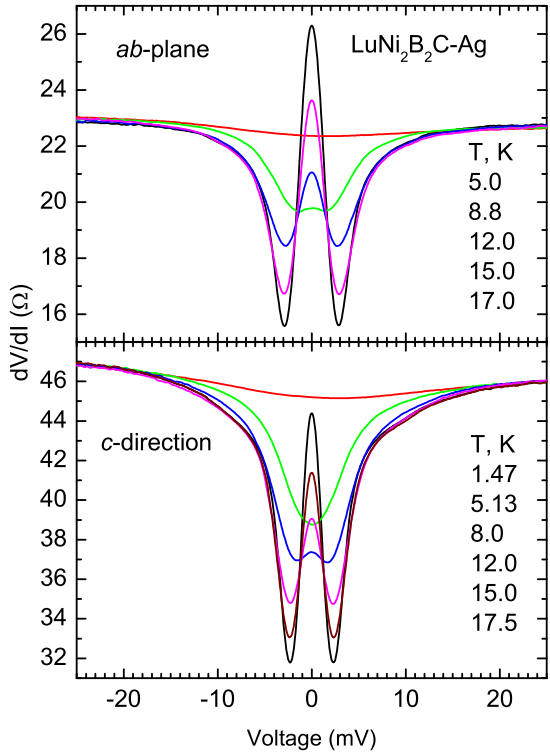}\\[12pt]
\caption[]{ Differential resistances of LuNi$_{2}$B$_{2}$C-Ag point contact at
different temperatures. To avoid crowding, only several curves are shown.}
\label{fig1}
\end{figure}

Some curves of the temperature series for the first derivatives $dV/dI$ of the
LuNi$_{2}$B$_{2}$C-Ag point contacts in the $c$ direction and in the $ab$ plane
are shown in Fig.\,\ref{fig1}.
 The contact diameters were calculated by Sharvin's equation using
$R_0^{ab}=22.5 \Omega$, $R_0^{c}=45 \Omega$, $\rho {l}=11.25\cdot
10^{-16} \Omega m^2$ of~\cite{Shulga} and allowing for the presence
of a potential barrier in the constriction region: $Z_{ab}=0.7$,
$Z_c=0.55$. The obtained diameters $d_{ab}=13.7 nm$ and $d_c=8.5 nm$
are very close in the order of magnitude to the coherence length in
this compound ($\xi=6.5 nm$~\cite{Shulga}). The measured curves were
symmetrized $dV/dI_{Sym}=1/2[dV/dI(V)+dV/dI(-V)]$ and normalized to
the normal state at $T > T_c$.

Two approaches were used to compare the theoretical and experimental
results obtained in the one- and two-gap approximation. First, a
model was applied, which describes the electrical conductivity of
pure $S-c-N$ point contacts in the presence of an arbitrarily
transparent potential barrier at the boundary between the metals.
The model allows for the finite lifetime of Cooper
pairs.~\cite{Beloborodko} The equations describing the $I-V$
characteristics of the point contact within this model are given
in.~\cite{Bobrov,Beloborodko} The model, however, interprets the
superconducting order parameter and the energy gap as distinct
magnitudes. The BCS order parameter $\Delta$ was found from

\begin{equation}
\Delta =\lambda \int \limits _{0}^{\omega_{D} }d\varepsilon
\tanh\left(\frac{\varepsilon }{2T} \right)Re\frac{1}{\sqrt{u^{2}
-1} }
\end{equation}
where $u$ is obtainable by solving the equation
\begin{equation}
\frac{\varepsilon }{\Delta } = u \left[ 1- \frac{\gamma }
{\sqrt{1-u^{2}} }  \right]
\end{equation}
The energy gap $\Delta_0$ and the order parameter $\Delta$ are
related as
\begin{equation}
 \Delta _{0} =\Delta \left(1-\gamma ^{2/3} \right)^{3/2}
\end{equation}
Here $\gamma=1/\left( \tau _s\Delta\right)$ is the pair breaking
parameter, $\tau_s$ is the mean free time during spin-flip
scattering at impurities. When magnetic impurities are absent,
$\tau_s$ tends to infinity and the equations describing the $I-V$
characteristics in~\cite{Beloborodko} and in~\cite{Blonder}
coincide.

The other approach was based on the generalized
Blonder-Tinkham-Klapwijk (BTK) model commonly used to describe
$S-c-N$ point contacts. The model allows for finite lifetime of
quasiparticles $\tau=\hbar /\Gamma$ determined by inelastic
scattering, which leads to the broadening of the density of states
in the superconductor~\cite{Plecenik}. Formally, according to the
previous theory, the BTK-based results are obtained under condition
of strong pair breaking $(|u| \gg 1)$~\cite{Beloborodko}. Therefore,
in the strict sense, the generalized BTK model contains no gap. For
any infinitesimal broadening parameter $\Gamma$, at $T=0$ the
density of states near Fermi surface is nonzero. In
theory~\cite{Beloborodko}, the order parameter is qualitatively
analogous to the pseudogap in the generalized BTK model.

In the two-gap approximation, the general conductivity of the point contact is
taken as a superposition of the conductivities from two regions of the Fermi
surface having their own gaps. In this case the experimental curves were
fitted using the expression

\begin{equation}
\frac{dV}{dI}=\frac{S}{\frac{dI}{dV}\left( \Delta _{1},\gamma
_{1},Z\right) K+\frac{dI}{dV}\left( \Delta _{2},\gamma
_{2},Z\right) \left( 1-K\right) } \label{two-gap}
\end{equation}

$K$ is the coefficient, characterizing the contribution from the
part of the Fermi surface with the gap $\Delta _1$, $S$ is the
scaling factor describing the intensity of the experimental curve.
It is used to make the amplitudes of the theoretical and
experimental curves equal. The best agreement between the shapes
of the theoretical and experimental curves corresponding to the
minimum $rms$ deviation in the curve $F(\Delta)$ was taken as the
main criterion of fitting in both the models at low and moderate
temperatures. When the temperature increases, the curves
$F(\Delta)$ become flatter, which entails much uncertainly in
estimation of $\Delta$. In this case the scale-factor $S$ is of
primary importance. It is, as a rule, independent of temperature
and at low temperatures it can be calculated quite accurately
through averaging. The details of the calculation in the one-gap
and two-gap approximations are offered in the Appendix.

\section{Calculation}

\subsection{One-gap approximation}

\begin{figure}
\centering
\includegraphics[width=\columnwidth]{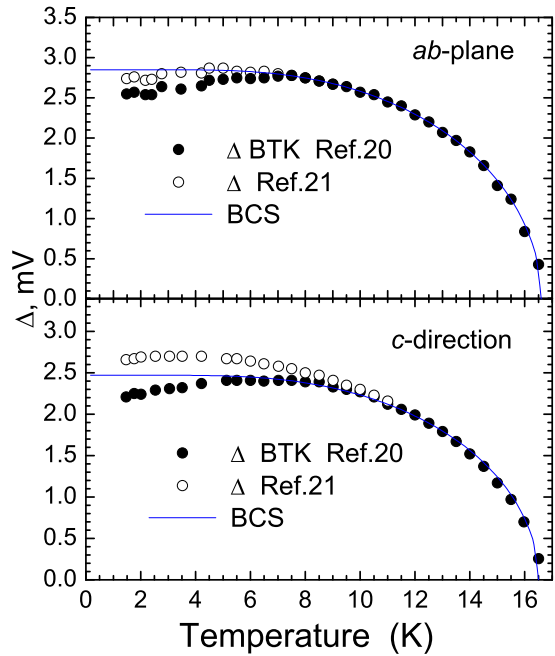}\\[12pt]
\caption[]{Temperature dependences of the order parameter (energy
gap) $\Delta$ calculated in the one-gap
approximation~\cite{Beloborodko} (open circles) and within the
generalized BTK model~\cite{Plecenik} (solid circles) for point
contact LuNi$_{2}$B$_{2}$C-Ag in the $ab$ plane and in the $c$
direction. Solid lines show the BCS approximation. Here and in all
other figures the following symbols are used: circles - one-gap or
average gap calculation; squares - the small gap; triangles - the
large gap; closed symbols - BTK calculation
Ref.\,\onlinecite{Plecenik}; open symbols - calculation by formulae
of Ref.\,\onlinecite{Beloborodko}} \label{fig2}
\end{figure}

Although LuNi$_{2}$B$_{2}$C exhibits a two-band kind of
superconductivity, its experimental curves can be described in the
zero approximation with an averaged gap possessing the broadening
parameter $\Gamma$ or the pair-breaking parameter $\gamma$. The
temperature dependences of the order parameters (the model
of~\cite{Beloborodko}) and the gaps (BTK
calculation~\cite{Plecenik}) in the $ab$ plane and in the $c$
direction are shown in Fig. \,\ref{fig2} along with the BCS curves.

In both cases we observe a deviation from the BCS curve. In the $c$
direction the deviation is towards higher values when the
calculation is based on~\cite{Beloborodko} and towards lower values
in the BTK approximation. As for the $ab$ plane, it should be noted
that in the low temperature region the resistance of the contact
varied during the measurement and became stable only at $4$ $K$.
This may be the reason why in both models the deviations from the
BCS dependence in the low temperature region descend to lower
values. The results of the both approximations start to coincide
near $8$ $K$ in the $ab$ plane and near $11$ $K$ in the $c$
direction. Precisely at these temperatures (see Fig. \,3), the
parameters $\gamma$ (pair breaking) or $\Gamma$ (broadening) turn to
zero in these orientations.

\begin{figure}
\centering
\includegraphics[width=\columnwidth]{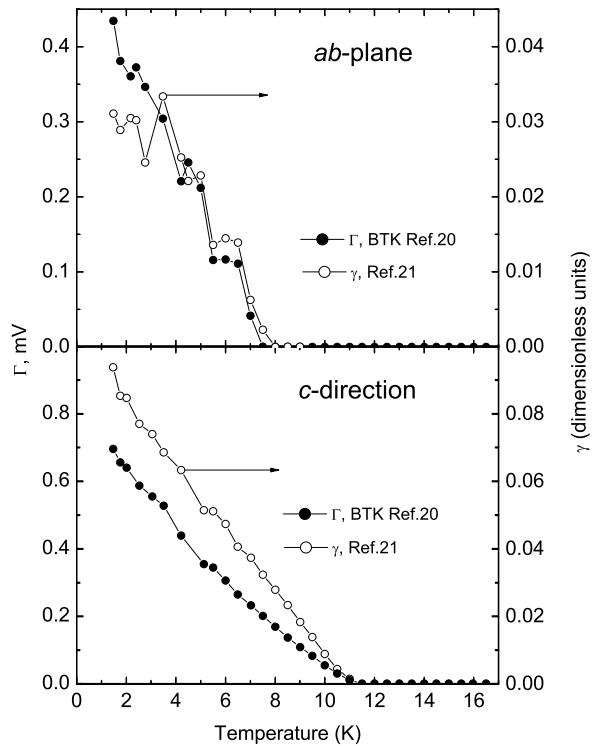}\\[12pt]
\caption[]{ Temperature dependences of broadening $\Gamma$ (closed
circles) and pair-breaking $\gamma$ (open circles) parameters for
LuNi$_{2}$B$_{2}$C-Ag point contact in the $ab$ plane and in the $c$
direction in the one-gap approximation.} \label{fig3}
\end{figure}

 The ratio $\Gamma/\Delta$ or the $\gamma$
magnitude can be used for semiquantitative estimation of the
contribution of the two-band structure to the point contact
conductivity. Since in the $c$ direction the ratio $\Gamma/\Delta$
is $1.5\div2$ times higher, we can expect a stronger two-band effect
in it. Besides, we can conclude that the experimental curves
demonstrate one-gap superconductivity above $8$ $K$ in the $ab$
plane and above $12$ $K$ in the $c$ direction. True, the more
accurate two-band calculation (see below) gives somewhat different
results; nevertheless, this technique is quite good for
semiquantative calculation. Note that in the strict sense, neither
$\Gamma$ nor $\gamma$ are meant for describing distribution of gaps
over the Fermi surface. They allow for the finite lifetime of
carriers, which only simulates the energy gap anisotropy. Their use
in this particular case is, however, quite justified as it enables
an approximate modeling of the $I-V$ characteristic in the $N-c-S$
point contact.

In $ab$ and $c$ orientations the gap is anisotropic, though observed
anisotropy is rather low (see Figs.\,\ref{fig1},\ref{fig2}).
Statistically (several tens of contacts were examined for each
direction), the positions of the $dV/dI$-minima characterizing the
gap value are $15\div20\%$ shifted towards higher energies in the
$ab$ plane.

\begin{figure}
\centering
\includegraphics[width=\columnwidth]{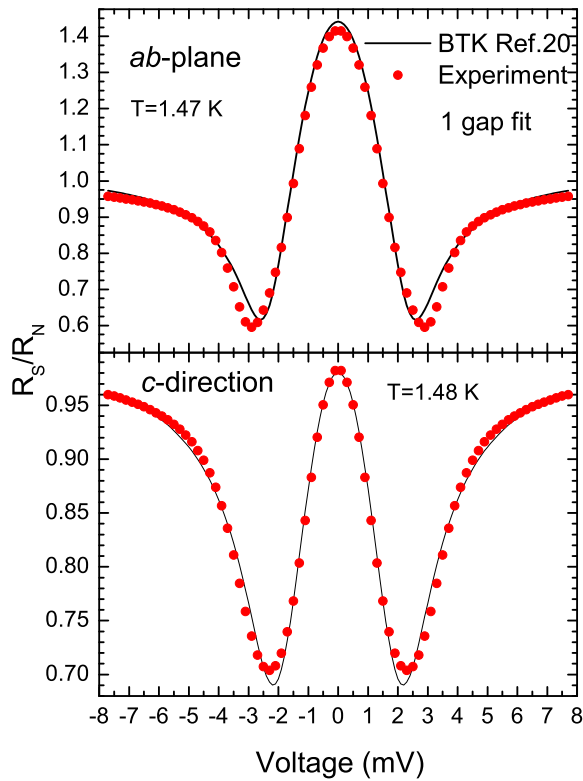}\\[12pt]
\caption[]{ Fitting of experimental curves in the one-gap
approximation (BTK model~\cite{Plecenik}) for LuNi$_{2}$B$_{2}$C-Ag
point contact. A few experimental points are shown to avoid
overloading.} \label{fig4}
\end{figure}

\begin{figure}
\centering
\includegraphics[width=\columnwidth]{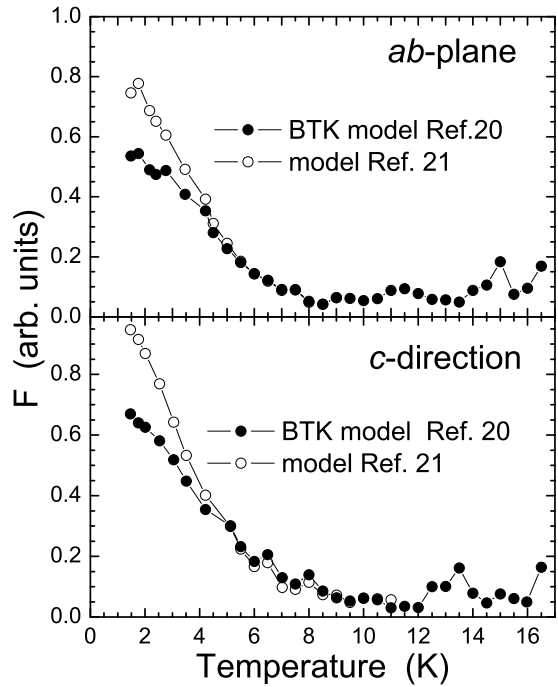}\\[12pt]
\caption[]{ Temperature dependences of the $rms$ deviation
characterizing shape discrepancies between experimental and
theoretical (calculation by Ref.\,\onlinecite{Beloborodko} curves
(open circles) and in the BTK model Ref.\,\onlinecite{Plecenik}
(closed circles)) for LuNi$_{2}$B$_{2}$C-Ag point contact in the
$ab$ plane (a) and $c$ direction (b) in the one-gap
approximation.} \label{fig5}
\end{figure}

In addition to the deviation from the BCS $\Delta(T)$-curve, the
one-gap model also fails to provide a good fitting of the shapes of
theoretical and experimental $dV/dI(V)$-curves. The one-gap fitting
(BTK model~\cite{Plecenik}) of experimental curves is shown in
Figs.\,\ref{fig4} and \ref{fig5} illustrates the temperature
dependences of the $rms$ deviations between the shapes of the
theoretical and experimental curves. The best agreement is observed
above $7 \div 8$~$K$.

 We attribute this rather high value to the two-band character of the
superconductivity. This indicates that experimental curves must be
fitted in the two-gap approximation, as in the case of MgB$_{2}$
and~\cite{Bobrov}.

\subsection{Two-gap approximation}

\begin{figure}
\centering
\includegraphics[width=\columnwidth]{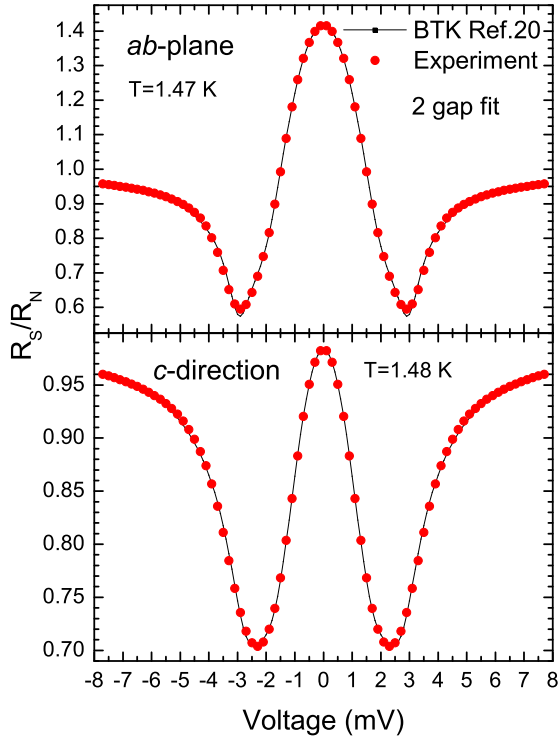}\\[12pt]
\caption[]{ Two-gap approximation (BTK model
Ref.\,\onlinecite{Plecenik} of experimental curves for
LuNi$_{2}$B$_{2}$C-Ag point contacts. To avoid crowding, only a
part of experimental points is shown.} \label{fig6}
\end{figure}

The two-gap fitting of experimental curves in the low temperature region is
illustrated in Fig. \,\ref{fig6}6. The agreement is seen to be much better than with the
one-gap approximation (compare with Fig. \,\ref{fig4}).

\begin{figure}
\centering
\includegraphics[width=\columnwidth]{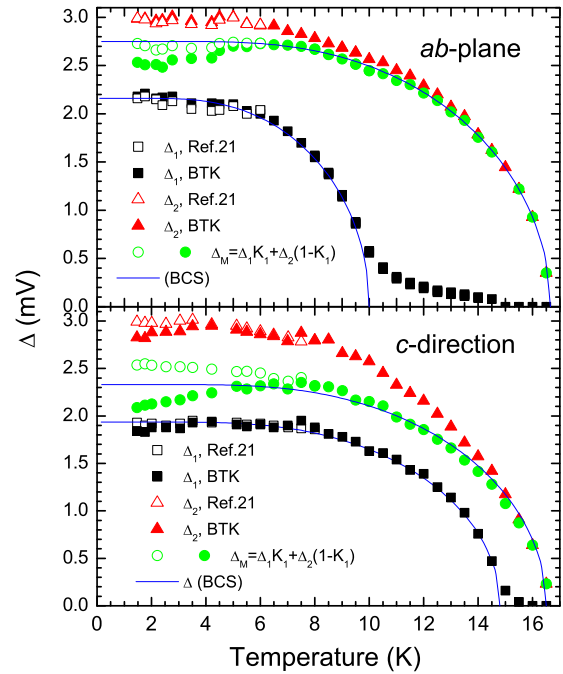}\\[12pt]
\caption[]{ Temperature dependences of the order parameter (energy
gap) $\Delta$ calculated in the two-gap approximation using the
equation of Ref.\,\onlinecite{Beloborodko} (open circles) and the
BTK model Ref.\,\onlinecite{Plecenik} (solid circles) for
LuNi$_{2}$B$_{2}$C-Ag point contacts in the $ab$ plane (a) and $c$
direction (b). Solid lines - BCS extrapolation. In addition to the
large $\Delta_2$ (triangles) and small $\Delta_1$ (squares) gaps,
the figure includes medium gaps $\Delta_M$ (circles) that were
obtained taking into account the partial contribution to the
contact conductivity from the Fermi surface regions with the large
and small gaps (see the formula in the figure and Eq.(4) in the
text).} \label{fig7}
\end{figure}

The temperature dependences of the order parameters calculated
from~\cite{Beloborodko} (open symbols) and the gaps calculated
within the generalized BTK model~\cite{Plecenik} (solid symbols) in
the $ab$ plane and along the $c$ axis are shown in Fig.\,\ref{fig7}
along with the corresponding BCS dependences (solid lines). In
addition to the small $\Delta_1$ (squares) and large $\Delta_2$
(triangles) gaps, Fig. \,\ref{fig7} illustrates for a comparison an
averaged gap $\Delta_M$ (circles) which allows for the partial
contributions to the conductivity of the contact made by the Fermi
surface areas with different gaps (see formula in Fig.
\,\ref{fig7}).

\begin{figure}
\centering
\includegraphics[width=\columnwidth]{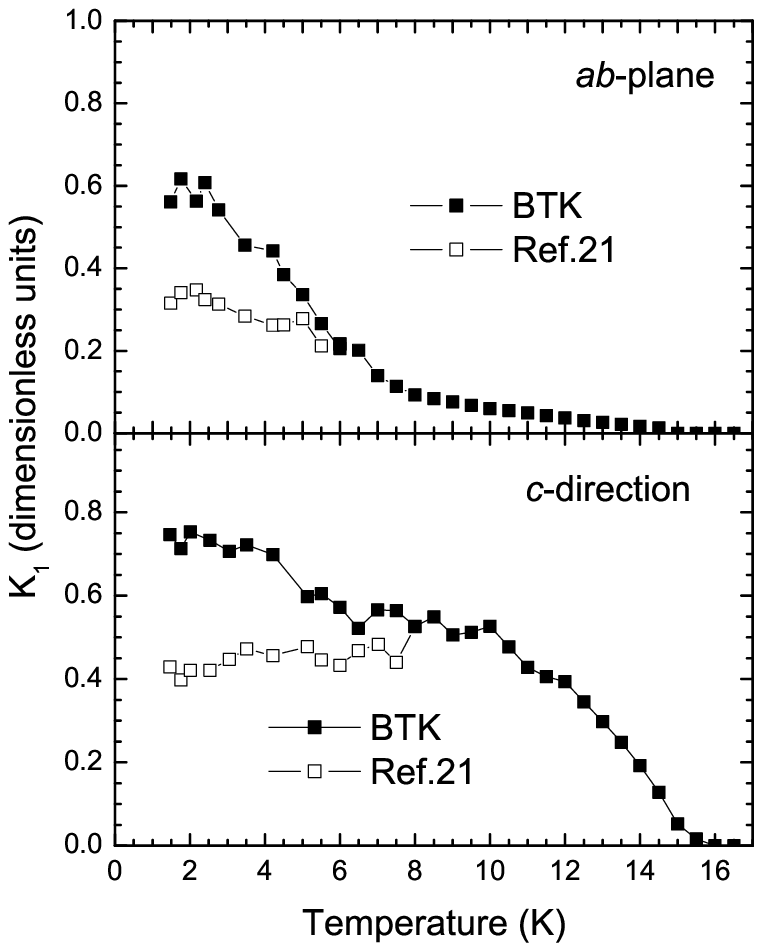}\\[12pt]
\caption[]{ Temperature dependences of the partial contribution to
the contact conductivity from the small gap (calculation by the
formula of Ref.\,\onlinecite{Beloborodko} and within the BTK model
Ref.\,\onlinecite{Plecenik}. Also, see Eq.4 in the text.}
\label{fig8}
\end{figure}

It is interesting to compare the BCS approximation of the critical temperatures
of the small gap in the $ab$ plane and in the $c$ direction:
 $T_c^{ab}(\Delta_1)=10$ $K$ and $T_c^c(\Delta_2)=14.6$ $K$. The absolute values of
the small gaps are close ($\Delta_0^{ab}=2.16$ $mV$,
$\Delta_0^c=1.94$ $mV$), the difference being no more than $10\%$.
The critical temperatures of the large gaps are similar and coincide
with the bulk $T_c=16.8$ $K$. The values of the large gaps are about
$\Delta_2=3$ $mV$. Note that both the models give practically
identical results for the large and the small gaps (open and closed
triangles and squares). As for the average gaps (circles), their
values in the low temperature region are dependent on the model
applied. This is because of different correlations between the
partial contributions from the large and the small gaps (see Fig.
\,\ref{fig8}).

It thus turns out that the higher $\Gamma$ (or $\gamma$) in the $c$
direction (where the two-band character is more pronounced) obtained
in one-gap approximation is determined not by the difference between
the values of the large and the small gaps but by the difference
between the partial contributions from these gaps. In the $ab$ plane
the contribution of the small gap (open squares, Fig. \,\ref{fig8}
upper panel) is smaller than in the $c$ direction and it decreases
rapidly as the temperature rises. In the $c$ direction the
contributions from both gaps are very close (in average between the
two models~\cite{Plecenik,Beloborodko} is about $50\%$ for each) up
to $10\div11$ $K$. (Note that in the BTK calculation~\cite{Plecenik}
the contribution from $\Delta_1$ is over $50\%$; in the calculation
by model of~\cite{Beloborodko} the $\Delta_1$ contribution is below
$50\%$ and in average the contribution from $\Delta_1$ is close to
that from $\Delta_2$ at low temperatures.)

\begin{figure}
\centering
\includegraphics[width=\columnwidth]{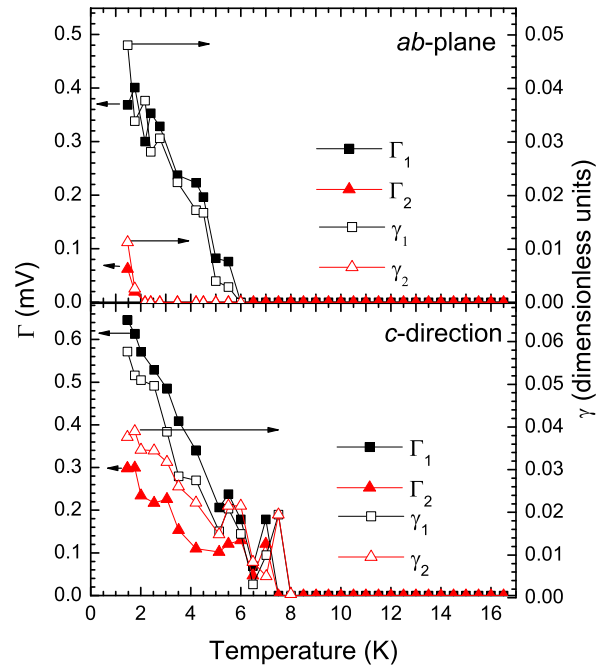}\\[12pt]
\caption[]{ Temperature dependences of the broadening $\Gamma$
(closed circles) and pair-breaking $\gamma$ (open circles)
parameters for LuNi$_{2}$B$_{2}$C-Ag point contact in the $ab$
plane (a) and in the $c$ direction (b) in the two-gap
approximation. Note that the parameters are considerably higher
for the small gaps (squares).} \label{fig9}
\end{figure}

The temperature dependence of $\gamma$ (pair breaking), or
$\Gamma$ (broadening of quasiparticle levels) obtained in the
two-gap approximation are shown in Fig. \,\ref{fig9}.

Note that the parameters $\Gamma$ and $\gamma$ are considerably
higher for the small gap both in the $ab$ plane and in the $c$
direction. Their ratios to the corresponding gaps are larger too:
$(\Gamma_1/\Delta_1)=0.185>(\Gamma_2/\Delta_2)=0.021$ in the $ab$
plane and $(\Gamma_1/\Delta_1)=0.333>(\Gamma_2/\Delta_2)=0.1$ in
the $c$ direction. As stated above, we consider that these ratios
simulate qualitatively to what extent the gap is distributed over
the corresponding sheet of the Fermi surface. We can thus assume
that at low temperatures the value of the gap changes very little
for the part of the Fermi surface with the larger gap
($\Delta_2$). In the small-gap region ($\Delta_1$), the gap is
spread over a considerably broader range of energies at low
temperatures.  Thus the two-gap approximation, even being more
accurate, cannot describe the experimental curves in the whole
$T$-interval. This is evident (see Fig. \,10) in the temperature
dependences of the $rms$ deviations $F$ between the shapes of the
fitted and experimental curves (even for two-gap calculation).

Although the errors observed (the $F$ magnitude) are much lower
(about half as high at the lowest temperature) than in the one-gap
case (Fig. \,\ref{fig5}), they increase appreciably below $T=4.5
\div 5$ $K$. This may occur because the three-band conductivity
comes into play, which must not be ruled out completely for this
compound. Proceeding from the significantly higher ratio
$\Gamma_1/\Delta_1$ at low temperatures, we can treat the small gap
in its turn as a superposition of two different gaps with different
$T_c$. These gaps are close to each other and their energies
overlap. The temperature dependence of the error ($F$) suggests that
we can expect $T_c \approx 5$ $K$ for the smallest gap in the BCS
approximation. This is the lowest estimate. However, because
$\Gamma_2$ becomes zero at $6$ $K$ ($a$ plane) and $8$ $K$ ($c$
direction) (Fig. \,\ref{fig9}), it is possible that precisely these
temperatures are closest to $T_c$ of the gap in the third band.

\begin{figure}
\centering
\includegraphics[width=\columnwidth]{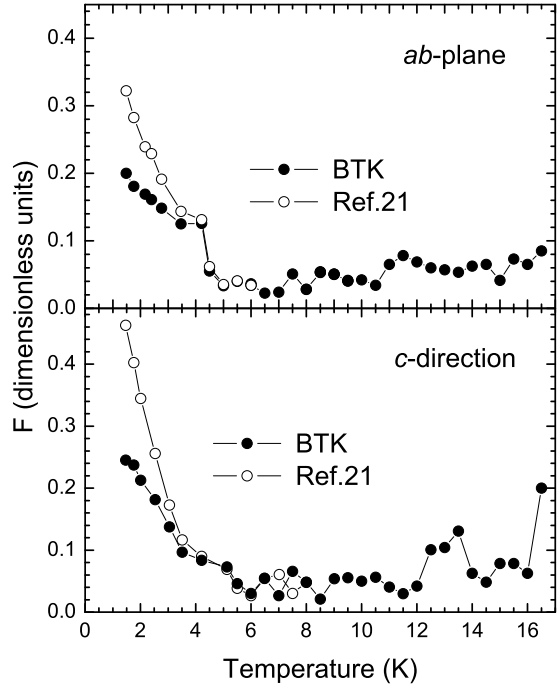}\\[12pt]
\caption[]{ Temperature dependences of the $rms$ deviation
characterizing shape discrepancies between experimental and
theoretical curves (calculation by the equations of
Ref.\,\onlinecite{Beloborodko} (open circles) and in the BTK model
Ref.\,\onlinecite{Plecenik} (solid circles)) for
LuNi$_{2}$B$_{2}$C-Ag point contacts in the $ab$ plane (a) and in
the $c$ direction (b) in the two-gap approximation.} \label{fig10}
\end{figure}

\section{Discussion}

In our experiments we have not succeeded in detecting noticeable
anisotropy of the energy gap in the $ab$ plane and in the $c$
direction. It was no more than 20\% in the average-gap
approximation. The reason may be a rough cleavage surface, which
made it difficult to fix the contact axis precisely along $a$
direction where the gap minimum is predicted. Therefore, following
the common practice in point contact spectroscopy, we tried to
select point contacts with the highest superconducting parameters
(gaps, $I-V$ nonlinearities). Since in the $ab$ plane the maximum
gap is predicted in the $\langle110\rangle$ direction, our contacts
probably were made to correspond most closely to this orientation.
Besides, the angular selectivity of point contact measurement is not
high enough. Since tunneling component in the point contacts is
rather weak (the $Z$ coefficient is not high) and their diameters
are close to the coherence length, it is quite likely that the
conversion of the electrons into pairs in the vicinity of the
contact occurs within a spherical geometry. According to Maki's
model~\cite{Maki} the angle between the largest and the smallest
gaps in the $ab$ plane is $45^\circ$. As a result, the spectra taken
in these directions can hardly have big distinctions. The most
obvious features discovered by us are the different critical
temperatures of the small gap in the $ab$ plane and in the $c$
direction obtained in the BCS extrapolation (Fig. \,\ref{fig7}).

With the evidence available, it is hardly possible to conclude unambiguously
whether this phenomenon is related to the gap anisotropy or any other (e.g.,
technological) factor. In the $ab$ plane the point contact was formed on a
freshly cleaved surface. In the $c$ direction the surface was cleaned
chemically and experienced no mechanical stress. It is known that elastic
scatterers first cause gap isotropization, then reduce the gap and suppress the
critical temperature. Although in our experiments the critical temperatures of
the point contacts coincide with $T_c$ of the bulk, the partial $T_c$ of the
small gap ($\sim 10 K$ Fig. \,\ref{fig7}) and its relative contribution (coefficient $K_1$
Fig. \,\ref{fig8}) to the general conductivity decreases in the $ab$ plane, which
indicates isotropization of the general gap in this plane. To clear up possible
reasons for this phenomenon, further measurements on chemically cleaned natural
growth faces of single crystals in the $a$ direction are needed.

According to Pokrovsky's theorem~\cite{Pokrovsky,Pokrovsky1} in the
case of anisotropic one-gap superconductivity, the temperature
dependence of the gap is independent of crystallographic directions.
In our case the temperature dependences of the large and the small
gaps and the BCS extrapolation of the critical temperatures are
quite different. This prompts the conclusion that this behavior can
not be caused by the contributions to conductivity made by a single
anisotropic superconducting gap in different crystallographic
directions. On the contrary, we observe the indications of two
different superconducting gaps from different energy bands.

Noteworthy also is another point, namely, the limits of
applicability of the two-gap approach to describing experimental
data. In the strict sense, the method requires that the broadening
(pair breaking) parameters should be zero for both gaps. This
condition is met at $T > 6$ $K$ for $ab$ plane and $T > 8$ $K$ in
the $c$ direction (see Fig. \,\ref{fig9}). At lower temperatures the
gaps do not have strictly specified values as they are distributed
over a certain range of energies simulated by nonzero $\Gamma$ or
$\gamma$. In the $ab$ plane the largest gap stands out as an
individual line above $2$ $K$, while in the $c$ direction both the
gaps start broadening at nearly the same temperature. As the
temperature lowers, the distribution extends up to overlapping of
energies. This brings more uncertainly into the description of
curves in the two-band approximation (Fig. \,\ref{fig10}) and sends
us in search of a new method. A possible technique of a more
adequate description of our experimental results near $T=1.5$ $K$ is
illustrated in~\cite{Muller} (Figs. \,{12-14}).

\section{Conclusions}

This study demonstrates that the two-gap approximation is more
appropriate for describing the superconductivity of
LuNi$_{2}$B$_{2}$C in a wide interval of temperatures. The values
and the temperature dependences of the large and the small gaps have
been estimated for $ab$ plane and the $c$ direction. It is found
that in the BCS extrapolation of the critical temperature for the
small gap is $T_c=10$ $K$ in the $ab$ plane and $T_c=14.5$ $K$ in
the $c$ direction. The absolute values of the gaps are
$\Delta_0^{ab}=2.16$ $meV$, $\Delta_0^c=1.94$ $meV$. For the large
gaps $T_c$ coincides with $T_c^{bulk}$: $T_c(\Delta_2)=16.8$ $K$ and
their absolute values are very close in both orientations, being
about $3$ $meV$. It is found that in the $c$ direction the
contributions to conductivity from the small and the large gaps are
nearly equal up to $10\div 11$ $K$. In the $ab$ plane the
contribution of the small gap is considerably weaker and decreases
rapidly as the temperature is growing.

\section*{Acknowledgment}

The single crystal samples for this study were graciously provided
by P.C. Canfield and S.L. Budko at Ames Laboratory and Iowa State
University. The authors are indebted to Yu.G. Naidyuk for helpful
discussions. The work was supported in terms of the complex program
of fundamental research "Nanosystems, nanomaterials and
nanotechnologies" of the National Academy of Sciences of Ukraine
($Project$ $No$ $10/05-N$). The work was supported in part by the
Robert A. Welch Foundation ($Grant$ $No$ $A-0514$, Houston, TX), The
Telecommunications and Informatics Task Force at Texas A\&M
University, the Texas Center for Superconductivity at the University
of Houston ($TCSUH$) and the National Science Foundation ($Grants$
$Nos.$ $DMR-010345$ and $DMR-0422949$). Partial support of U.S.
Civilian Independent States of the Former Soviet Union (Contract
No.\,UP1-2566-KH-03) is acknowledged.

\appendix{}
\section{}
 The iteration method commonly used to fit theoretical and
experimental curves is quite good when broadening is small. However,
as the broadening $\Gamma$ becomes comparable with the gap (or
$\gamma> 0.3$), this introduces an appreciable uncertainly into the
results. The iteration method implies that during fitting the
parameters change cyclically as the step is gradually decreased. It
is expected that the error reduces with each step. Our experiment
shows that with high $\Gamma$ (or $\gamma$) this procedure brings us
finally to a local minimum point which is dependent on starting
$\Delta$, $\Gamma$ ($\gamma$) and $Z$ values and cannot correspond
to the global minimum. Moreover, this method is not valid for the
two-gap calculation. In our fitting procedure we therefore used the
technique of coordinate descent with a postponed
solution~\cite{Bobrov}. First, we specified an interval in which
$\Delta$ is searched for at a given temperature. The interval was
then subdivided into equidistant parts $\Delta_1, \Delta_2, ...,
\Delta_n$. Then $\Gamma$ ($\gamma$) and $Z$ were fitted for each
$\Delta_i$. The method used in~\cite{Bobrov} does not imply that the
error should decrease immediately after each $\Gamma$ (or $\gamma$)
step. The errors are compared after each $\Gamma$ (or $\gamma$) step
only when the $Z$-fitting is completed (this accounts for the term
"postponed solution"). In the programming language, the calculation
by this method reduces to embedded cycles. In the one-gap model, the
$Z$-fitting is an inner cycle and the $\Gamma$ (or $\gamma$) fitting
is an outer cycle. In the two-band calculation, the $Z$-fitting is
the innermost cycle, the fittings in $\Gamma_1$ and $\Gamma_2$
become outer and, finally, the $\Delta_2$-fitting is the outermost
cycle. This method provides an unambiguous solution independed of
the starting parameters. In the process of calculation the
intensities of the theoretical and experimental curves were
equalized and aligned over the ordinate before each step of
calculation of the average $rms$ deviation $F(\Delta_i)$. To do
this, the $y$-coordinates of the points in the theoretical curve
were multiplied by the scale factor $S$ (the procedure of making the
amplitudes of the theoretical and experimental curves equal). The
curve was then shifted along the $y$-axis by amount $B$. The values
of $S$ and $B$ were found from the condition of the minimum
$F(\Delta_i)$. The standard algorithm for determination of these
coefficients known as the least-square method is considered, for
instance, in~\cite{Kahaner}. As a result, for each $\Delta_i$ we
could obtain $\Gamma_i$ ($\gamma_i$) and $Z$ at which the difference
between the shapes of the theoretical and experimental curves
characterized by the $rms$ deviation $F(\Delta_i)$ was the smallest
one. The kind of calculation for different temperatures is
illustrated in~\cite{Bobrov} (Fig.\,3). In the two-gap case, the
$F(\Delta_i)$-curves are more flattened, and the curve has no
minimum at a temperature lower than in the one-gap case. However,
before this happens, we are able, as a rule, to find quite
accurately the scale-factor $S$ which is independent of temperature
but is strongly dependent on $\Delta$ (~\cite{Bobrov}, Fig.4). The
value of $S$ can vary from contact to contact but it is invariant
for a particular contact. To put it more accurately, at the values
of $\Gamma \sim 0$, $S$ is practically invariable for the minima in
the curves $F(\Delta)$ (see Fig. 9 in~\cite{Bobrov}). By choosing
$\Delta$-values which correspond to the determined magnitudes of
$S$, we can plot quite accurately the curve $\Delta(T)$ in the
high-temperature region.

\end{document}